\documentclass[12pt,epsf]{article}

\usepackage{amsmath,amssymb,graphicx}

\newcommand{\beq}{\begin{eqnarray}}% can be used as {equation} or  {eqnarray}
\newcommand{\eeq}{\end{eqnarray}}

\newcommand{\be}{\begin{equation}}
\newcommand{\ee}{\end{equation}}
\newcommand{\bear}{\begin{eqnarray}}
\newcommand{\eear}{\end{eqnarray}}
\newcommand{\Frac}[2]{\frac{\displaystyle #1}{\displaystyle #2}}

\title{
{\huge \bf  Higgs-Z-photon Coupling  from Effect of  Composite Resonances}
\vspace*{0.8cm}
 }

\author{Haiying Cai   \\
 \\ \normalsize\emph{Department of Physics, Peking University, Beijing 100871, China}  \\ }

\date{}   %{\today}
\begin{document}
\setcounter{page}{0} \maketitle
\thispagestyle{empty}
\vspace*{0.5cm} \maketitle  %\begin{center}

\begin{abstract}
We explore the Higgs-Z-photon coupling in the Minimal Composite Higgs Model with vector and axial resonances. The electroweak precision
measurement, i.e. S and T,  is estimated for this model.  We calculate the signal strength for Higgs decay into  Z-photon  and  notable enhancement is found in certain EWPT allowed  parameter region.
\end{abstract}

\thispagestyle{empty}

\newpage

\setcounter{page}{1}

\section{Introduction}
 The LHC experiment groups have announced their discovery for one scalar resonance at $126 ~ \mbox{GeV}$, with its properties being  relatively close to the long awaited Standard Model (SM) Higgs boson~\cite{Higgs}. In the SM, the Electroweak Symmetry Breaking (EWSB) is elegantly triggered  by  one fundamental scalar  field  with  hand putting  quadratic term and quartic term which develops  VEVs at the  $\mathcal{O} (100 ~\mbox{GeV}) $  scale.  The Higgs coupling to gauge bosons  is  proportional to  the mass  squared  with overall  scaling factor set to be  one in the SM,  and  significant deviation from  this value would indicate that new degrees of freedom appearing  in the particle spectrum is essential  for unitarizing the  scattering amplitude of longitudinally polarized gauge bosons. Since the mass of a fundamental Higgs  is quadratically sensitive to the cut off scale $\Lambda$, and inevitably  requires severe amount of fine tunings, it is very hard for us to believe that the SM is valid till the Planck Scale.  Motivated by the  ``naturalness''  problem,  people  have explored many possibilities for the extension of SM with alternative mechanism to realize the EWSB, among which a light composite Higgs emerging from a strong dynamics at the TeV energy scale is a rather plausible scenario~\cite{CHM}.  It is widely noticed that  one common feature in composite Higgs model is the modified Higgs coupling with its deviation occurring at the order of $v^2/f^2$ and further complicated by the mechanism of  partial compositeness,  therefore we intend to study whether the phenomenology signature in this type of model is  preferred by  the best fit of most recent LHC data.

The ATLAS and CMS collaborations reported the measurement in
several Higgs decay modes: $W W^* $, $Z Z^* $,  $ b \bar b $ and $
\tau \tau $ and $ \gamma \gamma$, presuming that the light Higgs boson
is SM like. Special attention is focused on an enhancement around
$1.5$-$2.0$ in the diphoton decay rate while the central values of
other decay modes are very close to  the SM expectation. More
statistics is crucial to determine whether this deviation is
merely from the system errors  or hints at the existence of new
physics. From  perspective of collider experiment, the loop
induced $h^0 \to Z (\ell^+ \ell )\gamma$ constitutes another
clean and reconstructable search channel due to the fact that  all the
final states are detectable~\cite{Gainer}.  Although  this
process  is not yet accessible  with current energy and
luminosity, it will be extensively probed in the near future LHC,
since the coupling of $Z$ gauge bosons is not universal and
determined by the isospin assignment for the specific particle. 
Precise measuring the signal strength in
each possible decay channel will provide  a unique opportunity  to
understand the property of Higgs particle and  reveal the nature
of  Electroweak Symmetry Breaking.

The major goal of this paper is to test the compatibility of
composite model  with current LHC measurement, and further discuss
the effects of  partial compositeness on $H$-$Z$-$\gamma$
vertex.  Our study in some sense follows  the literature,  where
people have explored possible corrections to this vertex  due to
extra scalars,  gauge bosons  and fermionic partners in certain
extensions of the Standard Model~\cite{Carena, Djouadi, Chiang}.
Obeying the electroweak symmetry, any charged particle which
contributes to  $H$-$\gamma$-$\gamma$ vertex through  the quantum
effect  will at the same time contribute to  $H$-$Z$-$\gamma$
vertex.  However in our model set up,  it is noticed that there
are  additional non-diagonal gauge interactions from mixing
effects which will exclusively contribute to a gauge invariant
amplitude for the latter process. The paper is organized as the
following. We first recall the main feature in a minimal composite
Higgs model and thereafter we are going to illustrate the
electroweak precision test using the confidence ellipse since the constraints
from the $S$ and $T$ parameters must be considered together. Furthermore the bound on
relevant parameters will be extracted. Finally we  present  the form factor for
$H$-$Z$-$\gamma$ vertex including new contribution from the
non-diagonal gauge interaction.

\section{ Composite Higgs  Model }
We continue to investigate the minimal scenario where the composite Higgs is realized as one pNGB from the $\mathcal{G}/\mathcal{H}$ coset space and we are interested in  exploring the truncated effective theory with the presence of the lowest level  resonances. Let us first review the basic model set up relying on the CCWZ prescription since the original global symmetry $\mathcal{G} $ is nonlinearly realized~\cite{CCWZ}. The global symmetry breaking pattern  is $SO(5) \to SO(4)$, as $SO(4) \simeq SU(2)_L \times SU(2)_R $, the custodial symmetry is preserved  and only  the subgroup $SU(2)_L \times U(1)_R$ is weakly gauged which results in  an explicit breaking of the global symmetry at tree level.  Obviously,  an extra  $U(1)_X$ is necessary to reproduce the right $U(1)_Y$ coupling strength, i.e.  $g_0^\prime  = g_0 g_X/ \sqrt{g_0^2 + g_X^2 } $ in the low energy scale  as well as accommodate the right hypercharge $Y =  T_R^3 + Q_X$  for fermions.

There are a finite number of goldstone fields counted as
$\mbox{dim} SO(5) - \mbox{dim} SO(4) = 4$, living in the coset
space of $SO(5)/SO(4)$, thus we are going to parameterize the
non-linear sigma field in the matrix form $U = \mbox{exp} (i \sqrt
2 \pi^{\hat a} T^{\hat a}/f)$,  with $f$ being the decay constant
for those pNGB fields. It is convenient to choose the unitary
gauge  since  only one composite Higgs will remain: $\pi^{\hat a}
= (0, 0, 0, h^0) , \hat a = 1,2,3,4 $, once we turn on the $SU(2)
\times U(1)$ gauge interactions. One can calculate the gauge
invariant CCWZ structure following the normal procedure:
\beq
&  i U^\dagger D_\mu U =
d_\mu^{\hat a} T^{\hat a}  + E_\mu^{a_L} T^{a_L} + E_\mu^{a_R}
T^{a_R}  &   \nonumber \\
& D_\mu = \partial_\mu -i g_0 W_\mu^a T_L^a - i g_0^\prime B_\mu T_R^3 &
\eeq
with  $T^{\hat a } \, , \hat a = 1,2,3,4$ being generators in the broken direction and $T^{a_L,a_R } \,, a_{L,R} = 1,2, 3$  being generators in the unbroken direction. At the leading order of the chiral expansion, $d_\mu^{\hat a}$ and $E_\mu^{a_L, a_R}$ are expressed as:
\beq
d_{\mu}  &=& - \frac{\sqrt 2 }{f} \partial_\mu  h^0 \, T^{\hat 4 }  + \left( \frac{h^0}{\sqrt 2 f} - \frac{(h^0)^3}{6 \sqrt 2 f^3}   \right)  \left( g_0 \tilde{W}_\mu ^a- g_0^\prime \tilde{B}_\mu \delta^{a 3}\right) \delta^{a \hat i} \, T^{\hat i}  + \cdots \,  \nonumber  \\
E_{\mu}
&=&  \left(g_0 \tilde{W}_\mu ^a T_L^a + g_0^\prime \tilde{B}_\mu T_R^3  \right)- \frac{(h^0)^2}{4 f^2} \left( g_0 \tilde{W}_\mu ^a- g_0^\prime \tilde{B}_\mu \delta^{a 3}\right) (T_L^a- T_R^a) + \cdots  \label{Emu}
\eeq
It follows that  $d_\mu$ transforms covariantly under the local symmetry group, while  $E_\mu$ transforms like a gauge field.
Using the field strength of the external gauge fields $F_{\mu \nu} = F_{\mu \nu}^{a_L} T^{a_L} + F_{\mu \nu}^{a_R} T^{a_R} $,   a useful covariant tensor $  f_{\mu \nu}   \to  h(g,\pi )  f_{\mu \nu }  h^\dagger(g, \pi)$ can be constructed:
\beq
f_{\mu \nu} &=& U^\dagger\left( F_{\mu \nu}^{a_L} T^{a_L} + F_{\mu \nu}^{a_R} T^{a_R} \right) U \, = \, f_{\mu \nu}^++ f_{\mu \nu}^-  \nonumber  \\
  f_{\mu \nu}^+ &=& f_{\mu \nu}^{a_L} T^{a_L} + f_{\mu \nu}^{a_R} T^{a_R} \, , ~~  f_{\mu \nu}^-\, = \, f_{\mu \nu}^{\hat a} T^{\hat a}   \eeq
and performing a little bit algebra, we arrive the following relations:
\beq
f_{\mu \nu}^+ &=& \frac{1}{2} U^\dagger F_{\mu \nu} U + \frac{1}{2} U F_{\mu \nu} U^\dagger  \nonumber \\
f_{\mu \nu}^- &=& \frac{1}{2} U^\dagger F_{\mu \nu} U - \frac{1}{2} U F_{\mu \nu} U^\dagger \label{fuv}
\eeq
The kinetic term for goldstone bosons is  depicted by the gauge invariant operator:
\beq
&\mathcal{L}_{2} = {f^2}/{4} ~ \mbox{Tr} d_\mu d^\mu = \Frac{1}{2} \partial_\mu \pi^a \partial_\mu \pi^a  +  \Frac{a}{v}~ h^0 ~ \partial_\mu \pi^a \partial_\mu \pi^a + \cdots   & \\
& a  = \cos \theta = (1 -  s_h^2)^{1/2} \,, ~~~   s_h  =   \sin  \left\langle h^0 \right\rangle/ {f} \equiv  \sin (v/f) &
\eeq
where the Higgs coupling with other three pion fields is always less than 1,  thus new vector degrees of freedom are necessary to unitarize the $\pi \pi$ scatterings according to the partial UV completion hypothesis.

In this paper, we are going to introduce one vector resonance
$\tilde{\rho}^{a_L, a_R}_\mu$ transforming as $(3,1)\oplus (1,3)$
and one axial  resonance $\tilde{a}_\mu^{\hat a}$  transforming as
$(2,2)$  under the $SU(2)_L \times SU(2)_R$  symmetry group.  The
Lagrangian for vector and  axial resonances could be summarized by
the following equations~\cite{Contino,Marzocca}:
\beq
\mathcal{L}_{\rho_L} &=&  - \frac{1}{4} \mbox{Tr} \left( \tilde{\rho} _{L,\mu \nu } \tilde{\rho}_L^{\mu \nu } \right) + \frac{ f_\rho^2}{2}  \mbox{Tr} {\left( g_{\rho} \tilde{\rho}_{L\mu} - E_\mu ^L \right)^2}   \label{lr} \\
 \mathcal{L}_{\rho_R}  &=&  - \frac{1}{4} \mbox{Tr} \left( \tilde{\rho}_{R,\mu \nu } \tilde{\rho}_R^{\mu \nu } \right) + \frac{f_\rho^2}{2} \mbox{Tr}  {\left( g_{\rho} \tilde{\rho}_{R\mu} - E_\mu ^R\right)^2}   \label{rr}  \\
\mathcal{L}_a  &=&  - \frac{1}{4} \mbox{Tr} \left( \tilde{a}_{\mu \nu } \tilde{a}^{\mu \nu } \right) + \frac{f_a^2}{2 \, \Delta^2}  \mbox{Tr} {\left( {g_a} {a_\mu } + \Delta ~ {d_\mu } \right)^2}   \nonumber \\
&+& \frac{ i }{8} \mbox{Tr} \left( \left[ \tilde{a}_\mu, \tilde{a}_\nu \right]  \cdot \left[ \, U  \left( g_0 \tilde{W}_{\mu \nu} + g_0^\prime \tilde{B}_{\mu \nu} \right) U^\dagger + U^\dagger \left( g_0 \tilde{W}_{\mu \nu}+g_0^\prime \tilde{B}_{\mu \nu} \right) U \, \right] \right) \, \label{ar}
\eeq
where the couplings $g_\rho , g_a$ are assumed to be equal and  mass parameters can be defined as: $m_\rho =  g_\rho f_\rho$, $ m_a = g_a f_a /\Delta$. The electroweak field strengthes are: $ \tilde{W}_{\mu \nu} = \partial_\mu \tilde{W}_\nu -\partial_\nu \tilde{W}_\mu - i g_0 \left[ \tilde{W}_\mu ,  \tilde{W}_\nu \right]$, $\tilde{B}_{\mu \nu} = \partial_\mu \tilde{B}_\nu -\partial_\nu \tilde{B}_\mu$.  For the  axial resonances, the notation in the kinetic term reads:
\beq
\tilde{a}_{\mu \nu } = \nabla _\mu \tilde{a}_\nu  - \nabla _\nu \tilde{a}_\mu \, ,  \quad  \nabla _\mu  = \partial _\mu  - i E_\mu  \nonumber
\eeq
The mass terms  will give rise to the mixing between  electroweak gauge bosons and composite resonances:
\beq
\mathcal{L}_{a, \rm{mix}} &=& \frac{ m_a^2 \xi^{1/2}  }{ g_a v} \, \frac{\Delta}{\sqrt 2}  \,  h^0\tilde a _{\mu }^{\hat i} (g_0 \tilde W_\mu ^{\hat i}  - g_0^\prime  \tilde B_\mu \, \delta^{\hat i 3} )   \label{amix}  \\
\mathcal{L}_{\rho, \rm{mix}} &=& \frac{ m_\rho^2 \xi}{2  g_\rho v}  h^0\tilde \rho _{L\mu }^a (g_0 \tilde W_\mu ^a - g_0^\prime \tilde{B}_\mu \delta^{a 3} )  -\frac{m_\rho^2 \xi }{2 g_\rho v}  h^0\tilde \rho _{R\mu }^a(g_0 \tilde W_\mu ^a - g_0^\prime \tilde{B}_\mu \delta^{a 3} ) \label{rmix}
\eeq
with $\xi = v^2/f^2 $,  and  $ a, \hat{i} = 1,2, 3 $.  Since the magnitude of  $m_a^2  \xi^{1/2}$ or $m_\rho^2 \xi $ is not too much small, Higgs couplings in this class of  scenarios probably extend far away from the value points in the SM.

To facilitate the future discussion, we  particularly collect  the trilinear and quartic  gauge interactions  relevant for the $Z \gamma$ final state.  Notice that  axial fields are distinct in self interacting  from those of  other vector fields,  which can be directly derived from the kinetic terms in  Eq.[\ref{lr}-\ref{ar}]:
\beq
&& \mathcal{L}_{3G}  \, = \,  g_0 \varepsilon ^{abc} \partial _\mu \tilde{W}_\nu ^a  \tilde{W}_\mu ^b \tilde{W}_\nu ^c + g_\rho  \varepsilon ^{abc} \left( \partial _\mu  \tilde{\rho} _{L\nu }^a \tilde{\rho}_{L\mu }^b \tilde{\rho}_{L\nu }^c  +  \partial _\mu  \tilde{\rho} _{R\nu }^a \tilde{\rho}_{R\mu }^b \tilde{\rho}_{R\nu }^c  \right)  \, \nonumber \\
&& + \frac{1}{2}{\varepsilon ^{abc}}{\partial _\mu } ( g_0 \tilde{W}_\nu ^a + g_0^\prime \tilde{B}_\nu \delta ^{a3} ) \tilde{a}_\mu^b \tilde{a}_\nu ^c + \frac{1}{2}{\varepsilon ^{abc}} ( g_0 \tilde{W}_\mu ^a + g_0^\prime \tilde{B}_\mu \delta ^{a3} ) \tilde{a}_\nu ^b \left(\partial _\mu \tilde{a}_\nu ^c -\partial_\nu \tilde{a}_\mu^c \right)  \,  \label{trilinear}
\eeq
\beq
&& \mathcal{L}_{4G} \, = \, - \frac{{g_0^2}}{4}{\varepsilon ^{abc}}{\varepsilon ^{efc}} \tilde{W}^a_\mu \tilde{W}^b_\nu \tilde{W}^e_\mu \tilde{ W}^f_\nu - \frac{{g_\rho ^2}}{4}{\varepsilon ^{abc}}{\varepsilon ^{efc}} \left( \tilde{\rho}_{L \mu }^a \tilde{\rho}_{L \nu }^b  \tilde{\rho}_{L \mu }^e  \tilde{\rho}_{L\nu }^f + \tilde{\rho}_{R\mu }^a  \tilde{\rho}_{R\nu }^b  \tilde{\rho}_{R\mu }^e  \tilde{\rho}_{R\nu }^f  \right)   \nonumber  \\
&&  \quad \quad  - \frac{1}{8}\left( (g_0 \tilde{W}_\mu ^a + g_0^\prime \tilde{B}_\mu \delta ^{3a} )  \tilde{a}_\nu ^b - ( g_0 \tilde{W}_\nu ^a + g_0^\prime \tilde{B}_\nu \delta ^{3a} ) \tilde{a}_\mu ^b \right)  ( g_0 \tilde{W}_\mu ^a + g_0^\prime \tilde{B}_\mu \delta ^{3a} ) \tilde{a}_\nu ^b   \,
\nonumber  \\
&&  - \frac{1}{8}\left( (g_0 \tilde{W}_\mu ^a + g_0^\prime \tilde{B}_\mu \delta ^{3a} )  \tilde{a}_\nu ^b - ( g_0 \tilde{W}_\nu ^a + g_0^\prime \tilde{B}_\nu \delta ^{3a} ) \tilde{a}_\mu ^b \right)  ( g_0 \tilde{W}_\mu ^b + g_0^\prime \tilde{B}_\mu \delta ^{3b} ) \tilde{a}_\nu ^a  + \cdots  \,   \label{quartic}
\eeq
One theoretical prediction  from  Eq.[\ref{amix}-\ref{quartic}] is that,  axial and vector resonances  could manifest themselves through the decay channels: $ W^+  Z$, $h^0 W^+$, $q \bar q'$ and $t \bar b$, etc, after conducting the proper rotation for gauge bosons. The  $t \bar b$ channel will not be more viable than the light jets channel unless the third generation quarks carry a substantial degree of compositeness. It is emphasized  in Ref. \cite{Csaki, Barducci},  that one of the important approaches to  probe the  signatures of heavy vector bosons is through the  Drell-Yan and vector-boson-fusion (VBF)  production
processes.  In our phenomenology model,  assuming that  all the  fermions are completely fundamental, the di-bosons final state should be particularly explored other than the leptonic  channel and the multi-jets channel as they  dominate in the branching ratios of both the heavy neutral and charged vector bosons. Suppose that composite resonances primarily interact with those light SM quarks due to their alignments with EW gauge bosons,  we are able to extract  certain bound for the allowed resonance masses  exploiting information encoded in the cross
sections. In such a case  the direct search would not impose a too much stringent bound, hence the stronger constraint on the parameter space is expected to arise out of the Electroweak Precision Tests discussed in Section $3$.  On the other hand,  we notice that the mixing in the gauge sector significantly affects the loop induced Higgs decays and we will mainly explore those indirect effects later in Section $4$.

\section{ Electroweak Precision Tests}
In this section, we are going to discuss the Electroeweak
Precision Tests (EWPT), and  especially on constraint from the oblique
parameters. Oblique corrections to the Standard Model are encoded
in the vacuum polarizations  of gauge bosons which are quite
sensitive to the new physics fields with non-vanishing electroweak
couplings. The vacuum polarizations of a gauge boson can be
expanded around the zero momentum.
\beq \Pi _{a,a'} (p^2 ) = \Pi _{a,a'} (0) + p^2 \Pi '_{a,a'} (0) +
\cdots ,  \eeq
and we just consider the terms up to the order of $p^2$. One
calculable UV resonance contribution is from tree level exchange
of  vector and  axial fields,  thus the form factors after
integrating out the composite fields are:
\begin{figure}[h]
\begin{center}
\includegraphics[angle=0,clip,width=8.0 cm]{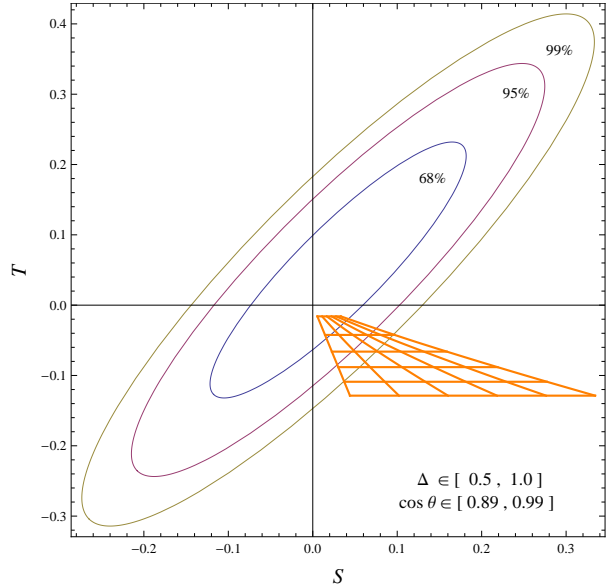}
\caption{{\small Contour plot for the S and T parameters  counting
the IR logarithmic contribution and UV  resonance effects at the
68\%  $(1 \sigma )$, 95\% $(2 \sigma )$ and 99\% $ (3 \sigma ) $
confidence levels. we adopt the benchmark point $g_a = g_\rho =
3.0$. }} \label{ST}
\end{center}
\end{figure}
\beq
\Pi '_{{W_3}B} (0) &=& - \left(\frac{s_h^2}{2 g_\rho ^2} - \frac{\Delta \, s_h^2}{2 g_a ^2}\right)  \, ,  \,   \Pi '_{W^ + W^ -}(0) = \Pi '_{W_3 W_3}(0) =  - \frac{g_0^2 + g_\rho ^2}{g_0^2g_\rho ^2} +  \left(\frac{s_h^2}{2g_\rho ^2} - \frac{\Delta \, s_h^2}{2g_a ^2}\right)  \nonumber  \\
 \Pi '_{B B} (0) &=& - \frac{g_0^{\prime 2} + g_\rho ^2}{g_0^{\prime 2} g_\rho ^2} + \left(\frac{s_h^2}{2g_\rho ^2} - \frac{\Delta \, s_h^2}{2g_a ^2}\right) \, , \quad     \Pi _{W^ + W^ -}(0) = \Pi _{W_3 W_3}(0) =   \frac{ f^2  }{4} s_h^2
\eeq
Note that  the additional identity  $\Pi _{B B}(0) =  -\Pi _{W_3 B}(0) =   \Pi _{W_3 W_3} (0) $ will ensure the electromagnetic gauge invariance.  In our analysis, we are going to adopt the notation in Ref.~\cite{peskin}, which are rescaled by $S = 4 s_w^2 \hat {S} / \alpha$ and $T = \hat{T} / \alpha$ compared with the notation employed in Ref.~\cite{Barbieri}.  Therefore  including the resonance contribution, the deviations of $S$, $T$ and $U$  parameters from their  SM  evaluations are:
\beq
\Delta S = - 16 \pi \cdot  \Pi '_{W_3 B} (0) = 8 \pi  \left(\frac{s_h^2}{ g_\rho ^2} - \frac{\Delta \, s_h^2}{ g_a ^2}\right) \, ,   \qquad  \Delta T = \Delta U =0  \,
\label{UV}\eeq
Considering  the gauge sector exclusively, another calculable source is the infrared (IR) contribution, which comes from the reduced gauge couplings with Higgs boson~\cite{Barbieri:2007}.
\beq
\Delta {S_{IR}} &=& \frac{1}{6 \pi} \left[ s_h^2 \log \left( \frac{\Lambda }{m_h} \right) + \log \left( \frac{m_h}{m_{h,ref}}\right)\right]   \,  ,    \\
\Delta {T_{IR}} &=&  - \frac{3  }{8 \pi c_w^2} \left[ s_h^2 \log \left( \frac{\Lambda }{m_h} \right)   + \log \left( \frac{m_h}{m_{h,ref}}\right) \right]
\label{IR} \eeq
where the effective cut-off  is defined as $ \Lambda = 4 \pi  f $.  Other UV contribution to $S$ and $T$ incorporates the effects at the one-loop level from higher dimensional operators in the chiral expansion, which is possible to make some tuning to the IR contribution given that it is not sensitive to the composite scale $\Lambda$~\cite{pich}. Therefore ambiguarity exists for the oblique parameters without UV completing the effective theory.  Extra calculable corrections to oblique parameters arise out of  vector-like composite fermions, whose influence can be analyzed using the equations in Ref.~\cite{Lavoura}.  At the one-loop level,  composite fermion contribution to the  $T$ parameter depends on the quantum number assignments in $SU(2) \times U(1)$, even though controled by its coupling to SM gauge bosons.  For singlet and doublet vector-like fermions, we expect to get substantially positive contribution to the $T$ parameter and compensate the negative contribution from the infrared  part  in Eq.[\ref{IR}]. Since the $S$  parameter is mostly positive under the condition $ 0< \Delta < 1$,  a positive $T$ parameter (from vector-like fermions) is  prefered to make the theory more compatible with the electroweak precison test.

As a rough estimate, we will just combine the results from Eq.[\ref{UV}- \ref{IR}] to interpret the bound imposed on relevant parameters.
The experimental values for $S$ and $T$ (leaving $U$ to be free) at $ 1 \sigma $  deviation are:
\beq
S = 0.03 \pm 0.10  \, ,  \quad   T = 0.05 \pm 0.12
\eeq
with the  correlation coefficient set to be $0.89$ and  the Higgs
reference mass  taken as $m_{h, ref} = 126.0 $ GeV. The  lines in
Fig. [\ref{ST}] show an important bound on the parameter space
$(\Delta,  \cos \theta)$ in this model from  present electroweak  precision  data.
For an input value  $\Delta = 0.5$,  it is
demanded  $\cos \theta \in (0.97, 1] $  to be consistent with
electroweak precision measurement at $99 \%$ C.L..  While  we
increase the value of $\Delta$, indicating more interplay between
electroweak gauge bosons and axial fields, the stringent
constraint from $S$ parameter will be relieved.  As we can see
with another input value $\Delta =1.0$,  a larger fraction of  region is
permitted, i.e.  $\cos \theta \in  (0.91, 1] $  at the $99 \%$ C.L.

\section{ Higgs-Z-photon coupling }
The model independent approach to discuss the Higgs phenomenology is adopting  Effective Field Theory, through parameterizing the couplings of Higgs bosons with gauge bosons and  fermions, deviations from new physics are accounted through the small corrections to the overall scaling factors \cite{Cacciapaglia}. Here we are going to employ the same effective Lagrangian used in the previous paper~\cite{cai}.
\beq \mathcal{L}_{eff} &=& a_W \frac{2m_W^2}{v} h^0 W_\mu ^+ W_\mu
^- + a_Z \frac{m_Z^2}{v} h^0 Z_\mu  Z_\mu  + c_\rho \frac{2
m_\rho^2}{v} h^0 \rho_{L \mu}^+ \rho_{L \mu}^-  + c_a \frac{2
m_a^2}{v} h^0 a_{ \mu}^+ a_{\mu}^-  \nonumber \\ &+&
 c_{a W}  \frac{ m_a^2}{v} h^0 \left( W_\mu^+ a_{\mu}^- +W_\mu^- a_{\mu}^+\right) +  c_{a \rho_L}  \frac{ m_a^2}{v} h^0 \left( \rho_{L \mu}^+ a_{\mu}^- +\rho_{L \mu}^- a_{\mu}^+\right) \nonumber \\ &+&  {c_f}  \left(
\frac{m_f}{v}\bar f f \right) h_0   +  c_\gamma ~ \frac{\alpha }{8 \pi
v } h^0 A^{\mu \nu } A_{\mu \nu }  + c_{ Z\gamma} \frac{\alpha }{4
\pi v }  h^0 Z^{\mu \nu } A_{\mu \nu } \eeq
After rotating into the mass eigenstates and keeping the terms only at the leading order of $g_0 / g_\rho$, $g_0^\prime / g_\rho $ and $\xi$, we get the  modified  Higgs couplings with various gauge bosons, i.e. $W_\mu^\pm, Z_\mu$, $a_\mu^{\hat i}$ and $\rho_\mu^{a_L, a_R}$ at the tree level:
\beq
a_{W} &=& \cos \theta +\frac{ g_0^2 \xi}{2  g_\rho^2} \frac{m_\rho^2}{m_W^2}  - \frac{ \Delta^2 g_0^2 \xi}{2 g_\rho^2} \frac{m_a^2}{m_W^2} \, \nonumber \\
c_a  &=&   \frac{\Delta^2 \xi g_0^2  \left( m_\rho^2 -m_a^2 \right) }{2  \left( g_0^2 m_\rho^2 + g_\rho^2 (m_\rho^2 -m_a^2)\right)} \, \nonumber \\
 c_{\rho } &=& -\frac{ g_0^2 \xi}{2 g_\rho^2} + \frac{\Delta^2 \xi g_0^4  m_a^4}{2 g_\rho^2 m_\rho^2  \left( g_0^2 m_\rho^2 + g_\rho^2 (m_\rho^2 -m_a^2)\right)} \, \nonumber \\
a_{Z} &=& \cos \theta+ \frac{\left(g_0^2 + g_0^{\prime 2} \right)
m_\rho^2 \xi }{ 2 g_\rho^2 m_Z^2}  - \frac{ \Delta^2 (g_0^2+ g_0^{\prime 2} ) m_a^2 \xi}{2 g_\rho^2 m_Z^2 }
\eeq
while the leading non-diagonal Higgs couplings appear at  the $\xi^{1/2}$ order:
\beq
c_{a W}  = \frac{\Delta}{\sqrt 2} \frac{ g_0 }{g_\rho} \xi^{1/2} \, , \qquad  c_{a \rho_L} = \frac{\Delta} {\sqrt 2} \frac{ g_0^2 }{ g_\rho^2} \xi^{1/2}
\eeq
Due to the mixing from Eq.[\ref{amix}-\ref{rmix}], non-diagonal  gauge couplings with one $a_\mu^\pm$ involved  occur at the $\xi^{1/2} $ order.  In fact there are also non-diagonal gauge couplings with one $\rho_L^\pm$ involved which will appear at the $\xi $ order after rotating into the mass basis is performed.  Hence the effects from the $\xi$ order vertexes  could be safely  ignored  since they only lead to higher order correction.  In the following discussion,  we take a further simplification, $m_a = m_\rho$, which would imply that there is no Higgs coupling to two axial fields, i.e. $c_a =0$, under such an assumption. Albeit  the mass of axial resonance is usually expected to be slightly  heavier than the mass of vector resonance, i.e. $ m_a > m_\rho$, we have checked that in such a situation the relevant phenomenology discussed in this section will not be altered  too much. It is worthwhile for us to show  that including non-diagonal couplings exclusively leads to a gauge-invariant contribution
for the form factor $c_{Z \gamma}^{(1)}$.
\begin{figure}[h]
\begin{center}
\includegraphics[angle=0,clip,width=14 cm]{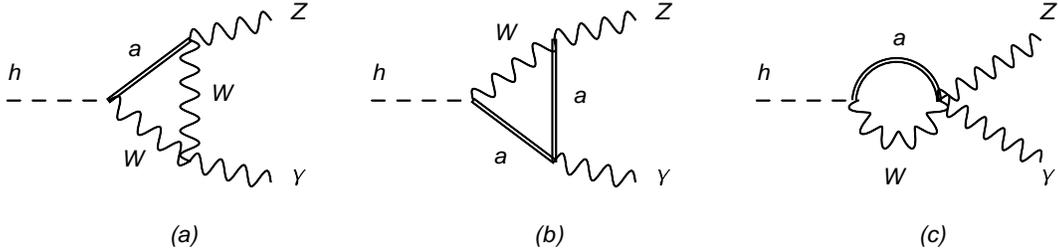}
\caption{{\small  (a) triangle feyndiagram with one axial resonance, (b) triangle feyndiagram with two axial resonances , (c) quartic feyndiagram with one axial resonance.}}
\label{feyn}
\end{center}
\end{figure}

The generic non-diagonal trilinear and quartic gauge self couplings can be put in the following way:
\beq
\mathcal{L}_{Z V_1 V_2} &=&  (- i  \, e \, \cot \theta_w ) \, c_{Z V_1 V_2} \bigg( \partial_\mu Z_\nu  V_{2, \mu}^+ V_{1, \nu}^- + \partial_\mu V_{2, \nu}^+ V_{1,\mu}^- Z_\nu + \partial_\mu V_{1, \nu}^- Z_\mu V_{2, \nu}^+   \nonumber \\ &+ &  \partial_\mu Z_\nu  V_{2, \nu}^- V_{1, \mu}^+ + \partial_\mu V_{2, \nu}^- V_{1, \nu}^+ Z_\mu + \partial_\mu V_{1, \nu}^+ Z_\nu V_{2, \mu}^-  - \left(\mu \leftrightarrow \nu \right)  \bigg)  \eeq
\beq
\mathcal{L}_{A Z V_1 V_2} &=&  (e^2 \, \cot \theta_w )  \, c_{Z V_1 V_2} \bigg(2 ~ Z_\mu A_\mu  V^+_{2, \nu} V^-_{1, \nu}  -Z_\mu A_\nu V^+_{2, \mu} V^-_{1, \nu}  -Z_\mu A_\nu V^+_{2 ,\nu} V^-_{1, \mu}  \nonumber  \\
&+ & 2 ~ Z_\mu A_\mu  V^-_{2,\nu} V^+_{1, \nu}  -Z_\mu A_\nu V^-_{2, \nu} V^+_{1, \mu}  -Z_\mu A_\nu V^-_{2, \mu} V^+_{1, \nu} \bigg)
\eeq
where the index exchange $ \left(\mu \leftrightarrow \nu \right)$ refers to the last two variables, and one  way to check the validity  is that the Lagrangian need to be self-conjugate.  Therefore in our model setup,  using the information provided in Eq.[\ref{trilinear}-\ref{quartic}],  at the leading order expansion, non-diagonal gauge couplings $c_{Z a W} $ and $ c_{Z a \rho_L} $ are expressed as:
\beq
c_{ZaW} = \frac{\Delta }{\sqrt 2}\frac{g_0^2 + g_0^{\prime 2}}{2 g_0 g_\rho }{\xi ^{1/2}}\, , \qquad   c_{Z a \rho_L } =  - \frac{\Delta }{\sqrt 2 }\frac{ \left(g_0^2 + g_0^{\prime 2} \right) }{2 g_0^2 } \, \xi ^{1/2}  \,
\eeq
To calculate the amplitude for $h^0 \to  Z \gamma$,  it is simply
to add up the three diagrams illustrated in Fig. [\ref{feyn}] and
times  a symmetry factor of two to account for another three
equivalent diagrams with the simultaneous external and internal
fields exchange: i.e. $ Z_\mu \leftrightarrow  A_\mu$ and
$W_\mu^\pm \leftrightarrow a_\mu^\pm $.  As we put all the
external particles to be on shell,  the amplitude can be organized
into  a gauge invariant form:
\beq
&& M({h^0} \to Z\gamma ) = 2 \cdot \left (M^{(a)}+ M^{(b)} + M^{(c)}  \right) \nonumber  \\
&=&\frac{ - i \, e^2}{8 \pi ^2 v} \cdot ( 2 ~ c_{V_1 V_2} \, c_{Z V_1 V_2} ) ~ c_{Z\gamma }^{(1)} (m_1, m_2) \, \left( {{g^{\mu \nu }}{k_1} \cdot {k_2} - k_1^\mu k_2^\nu } \right) \, \varepsilon _\mu ^*({k_1})\varepsilon _\nu ^*({k_2})
\eeq
Putting everything together, the form factor for $H$-$Z$-$\gamma$ vertex contains the  dominating contribution from both the fermion sector and the gauge sector:
\beq
 c_{Z \gamma} &=& c_t  \, N_c \, \frac{ (Q_t- 4 Q_t^2  s_w^2) }{s_w c_w} \, m_t^2 \, c_{Z \gamma}^f  \,(m_t) + a_W  \, c_{Z \gamma}^{(0)} \,(m_W)  \nonumber \\ &+&  \left( c_{\rho} +2\, c_{a \rho_L} \, c_{Z a \rho_L}  \right) c_{Z \gamma}^{(0)}{(m_a)} + 2\, c_{a W} \, c_{Z a W} \, c_{Z \gamma}^{(1)} \, (m_W, m_a)
\eeq
with each piece of  $c_{Z \gamma}$  being expressed by the one-loop  scalar and vector three point functions $C_0$ and $C_2$ defined in \cite{tHooft}:
\beq
 c_{Z \gamma}^{f} \,(m_t) &=& 4~  C_2 (m_h^2, m_Z^2,0,  m_t^2, m_t^2, m_t^2)  +  C_0 \left(m_h^2, m_Z^2, 0, m_t^2 ,m_t^2
,m_t^2 \right)
\eeq
\beq
 c_{Z \gamma}^{(0)} \, (m_1) &=& \cot \theta_w \cdot \bigg[ 2 \left(  \frac{m_h^2}{m_1^2} \,\left( {m_Z^2 - 2 m_1^2 } \right)
+ 2\, ( m_Z^2 -  6 m_1^2 )   \right)   C_2 (m_h^2, m_Z^2,0,  m_1^2, m_1^2, m_1^2)  \nonumber \\ &+&  4 \left( m_Z^2 - 4 m_1^2
\right) \cdot  C_0 \left(m_h^2, m_Z^2, 0, m_1^2 ,m_1^2
,m_1^2 \right) \bigg]
\eeq
\beq
c_{Z \gamma}^{(1)} \, (m_1, m_2) &=& \cot \theta_w \cdot \frac{m_2^2}{ m_1^2} \cdot  \bigg[ \left(  \left( {\frac{m_h^2}{m_2^2} + \frac{m_1 ^2}{m_2^2}  + 1 } \right) \left(m_Z^2 -m_2^2 -m_1^2 \right) -8 m_1^2 \right)   \nonumber \\
&\cdot &  \bigg( C_2 \left(m_h^2, m_Z^2,0,  m_1^2, m_2^2, m_1^2 \right) +C_2 \left(m_h^2, m_Z^2,0,  m_2^2, m_1^2, m_2^2 \right) \bigg)   \nonumber \\
 &+&   2 \left( \frac{m_1^2}{m_2^2} \left( m_Z^2 - m_1^2 - 3 m_2^2
\right) \right) \cdot C_0 \left(m_h^2, m_Z^2,0, m_1^2 ,m_2 ^2 , m_1^2
\right)  \nonumber \\
&+&   2 \left( m_Z^2 - 3 m_1^2 -m_2^2
\right) \cdot  C_0 \left(m_h^2, m_Z^2, 0, m_2^2 ,m_1^2
,m_2^2 \right) \bigg]
\eeq
where the $C_2$ function in the expression of $c_{Z \gamma}^{(0)} $ and the combination of  $C_2$ functions with different masses in the expression of $c_{Z \gamma}^{(1)}$ can be recasted into  Passarino-Veltman functions  $B_0$ and $C_0$:
\beq && C_2 \left(m_h^2, m_Z^2,0,  m_1^2, m_1^2, m_1^2 \right)  \nonumber \\ &=&  \frac{  m_Z^2 }{2 \left( {m_Z^2  - m_h^2 }
\right)^2}   \bigg( B_0 \left( {m_Z^2 ,m_1^2 ,m_1^2 } \right) - B_0 \left( {m_h^2 ,m_1^2 ,m_1^2 } \right)  \bigg)    \nonumber \\
&+&    \frac{1}{2 \left( {m_Z^2  - m_h^2 } \right)}  + \frac{m^2}{\left( m_Z^2-m_h^2 \right)} C_0\left( m_h^2 , m_Z^2,0 ,m_1^2, m_1^2 , m_1^2 \right)
\eeq
\beq && C_2 \left(m_h^2, m_Z^2,0,  m_1^2, m_2^2, m_1^2 \right) +C_2 \left(m_h^2, m_Z^2,0,  m_2^2, m_1^2, m_2^2 \right) \nonumber \\
&=&   \frac{ m_Z^2 }{\left( {m_Z^2  - m_h^2 }
\right)^2}   \bigg( B_0 \left( {m_Z^2 ,m_2 ^2 ,m_1^2 } \right) - B_0 \left( {m_h^2 ,m_2^2 ,m_1^2 } \right)  \bigg)    \nonumber \\
&+&    \frac{1}{\left( {m_Z^2  - m_h^2 } \right)}  + \frac{m_1^2}{\left( m_Z^2-m_h^2 \right)} C_0\left( m_h^2 , m_Z^2,0 ,m_1^2, m_2^2 , m_1^2 \right)  \nonumber \\ & +& \frac{m_2^2}{\left(m_Z^2-m_h^2\right)}  C_0 \left(m_h^2, m_Z^2, 0, m_2^2 ,m_1^2
,m_2^2 \right)
\eeq
Notice that in the limit of $m_2 = m_1$, our new form factor  $c_{Z \gamma}^{(1)}$ will reduce exactly to the $W$ gauge bosons mediated SM contribution $c_{Z \gamma}^{(0)}$~\cite{hza}.  Since the mass of axial resonance $m_a$ is much larger than the mass in $(m_h,  m_Z, m_W)$, it is meaningful to take the heavy mass limit expansion for  $c_{Z \gamma}^{(0)} (m_a) $ and $c_{Z \gamma}^{(1)}(m_W, m_a)$, which , to the leading orders,  could be asymptoted by the following approximations:
\beq
c_{Z \gamma}^{(0)} (m_a) & \simeq &  \cot  \theta_w   \cdot  \bigg(7 +\frac{\left(11 m_h^2-37 m_Z^2\right)}{30 m_a^2}\bigg) + \mathcal{O} \left(\frac{1}{m_a^4} \right) \nonumber \\
c_{Z \gamma}^{(1)} (m_W, m_a) &\simeq &  \cot \theta_w  \cdot \bigg(\frac{ 7 m_a^2}{4 m_W^2} + \frac{9}{2} \log \left(\frac{m_a^2}{m_W^2} \right) +  \frac{5 m_h^2-47 m_Z^2 -45 m_W^2}{36 m_W^2}  \bigg)   \nonumber \\  &+& \mathcal{O} \left(\frac{1}{m_a^2} \log \left(\frac{m_a^2}{m_W^2}\right) \right)
\eeq
%  & + &  \frac{5 m_h^2-47 m_Z^2 -45  m_W^2}{36 m_W^2}  \bigg) +\mathcal{O} (1/m_a^2)
It should be mentioned that  the non-diagonal contribution mildly makes up around $5 \%$-$10 \%$  contribution to the total $c_{Z \gamma}$ form factor provided that $m_a $ is around the TeV scale  and it interferes with the SM portion constructively.  Accordingly we are going to conclude that substantial correction is induced by the shift in $a_W$ after we take into account the mixing effects.
\begin{figure}[h]
\begin{center}
\includegraphics[angle=0,clip,width=7.5cm]{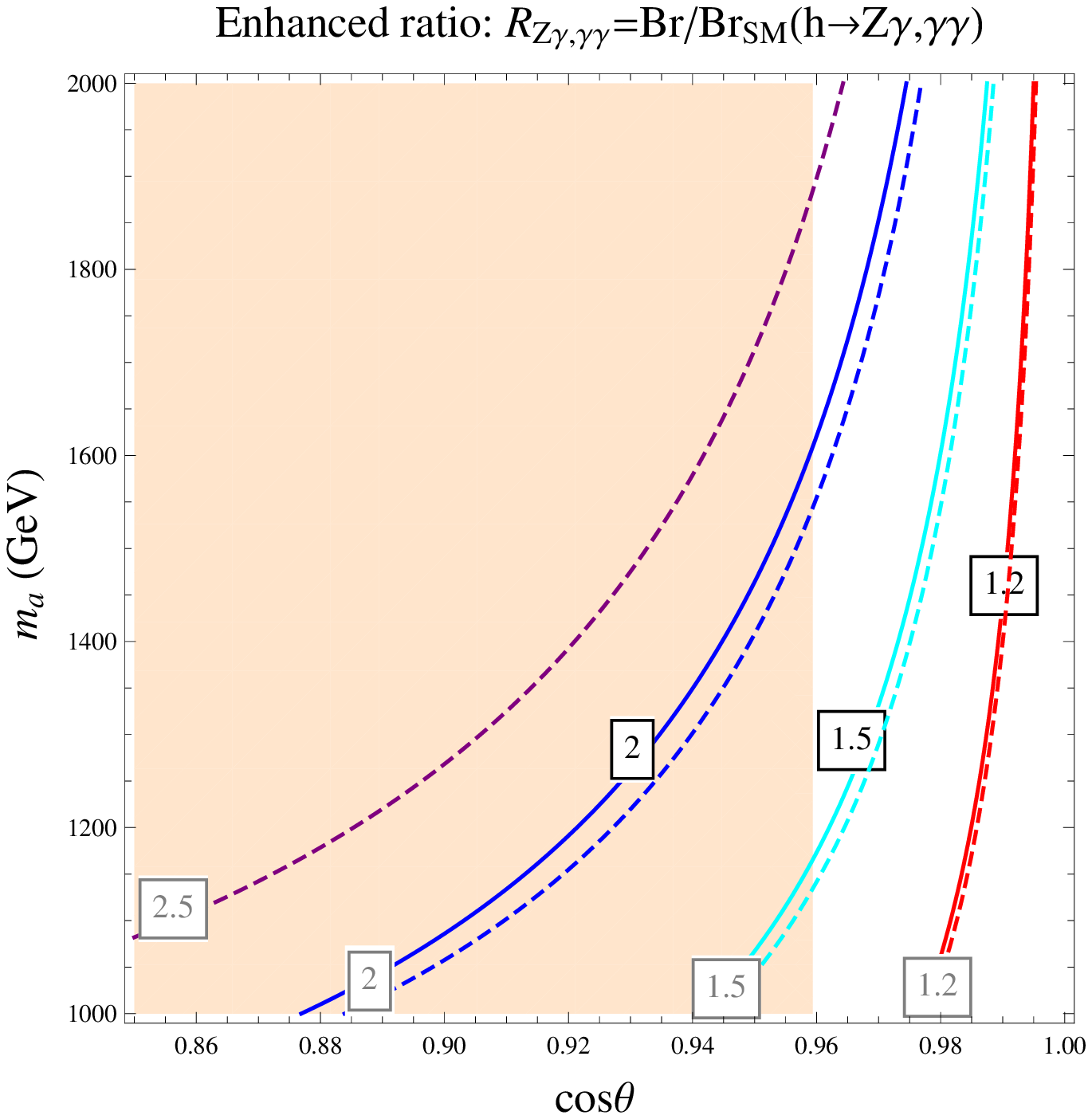}
\includegraphics[angle=0,clip,width=7.5cm]{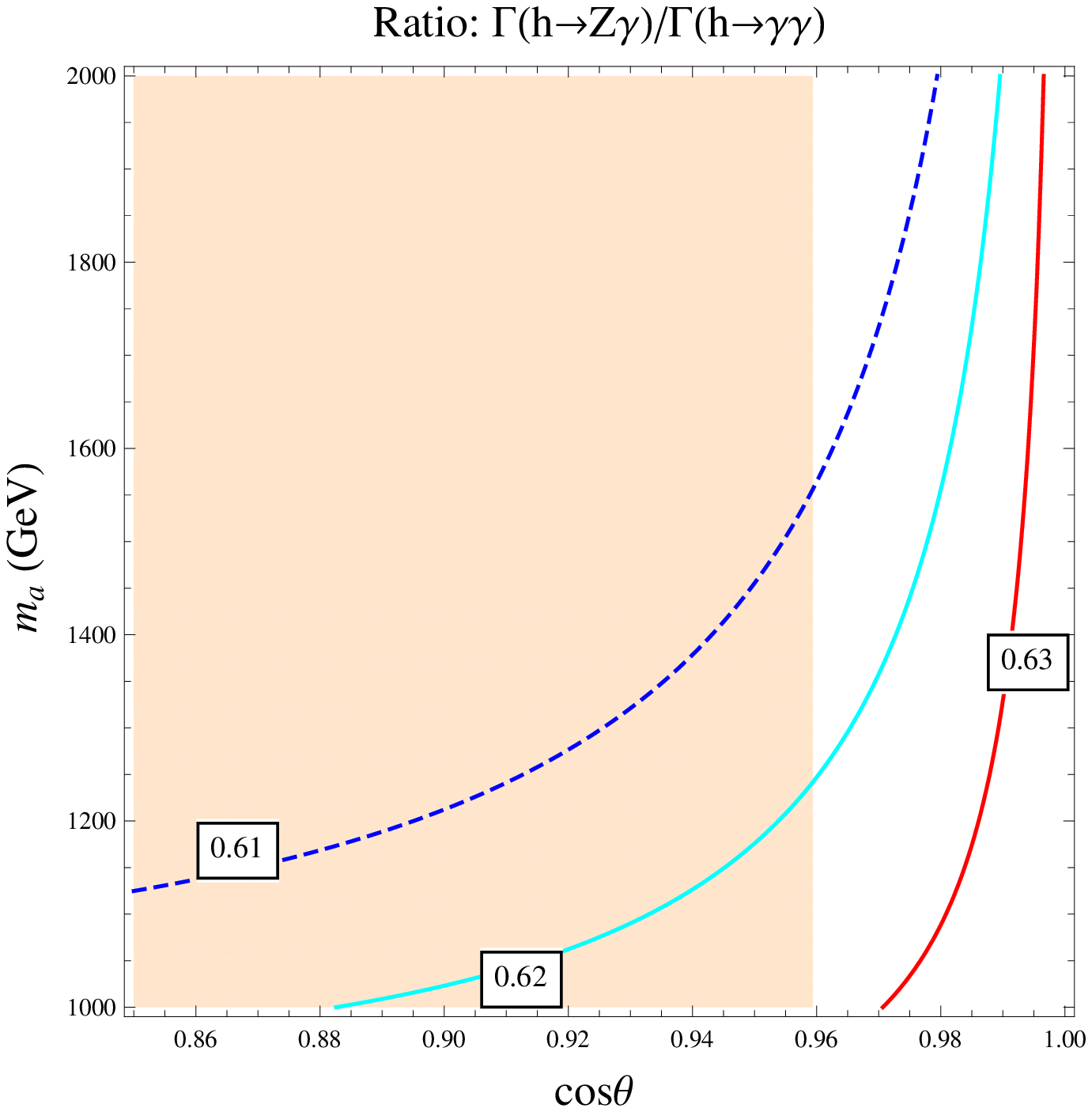}
\caption{{\small Left Contour:  signal strengths for $Z \gamma$ mode(solid line) and $\gamma \gamma$ mode (dashed line) in the $(m_a, \cos \theta )$ plane. The purple dashed line shows the upper bound measured by  ATLAS at $68 \%$ confidence level.  Right Contour: the decay width ratio between  $Z \gamma $ mode and $ \gamma \gamma$ mode in the $(m_a, \cos \theta )$ plane. The condition $\Delta = 0.5$  is imposed.  The light orange region is excluded by the EWPT requirement $S < 0.13$. }}
\label{ratio}
\end{center}
\end{figure}
\begin{figure}[h]
\begin{center}
\includegraphics[angle=0,clip,width=7.5cm]{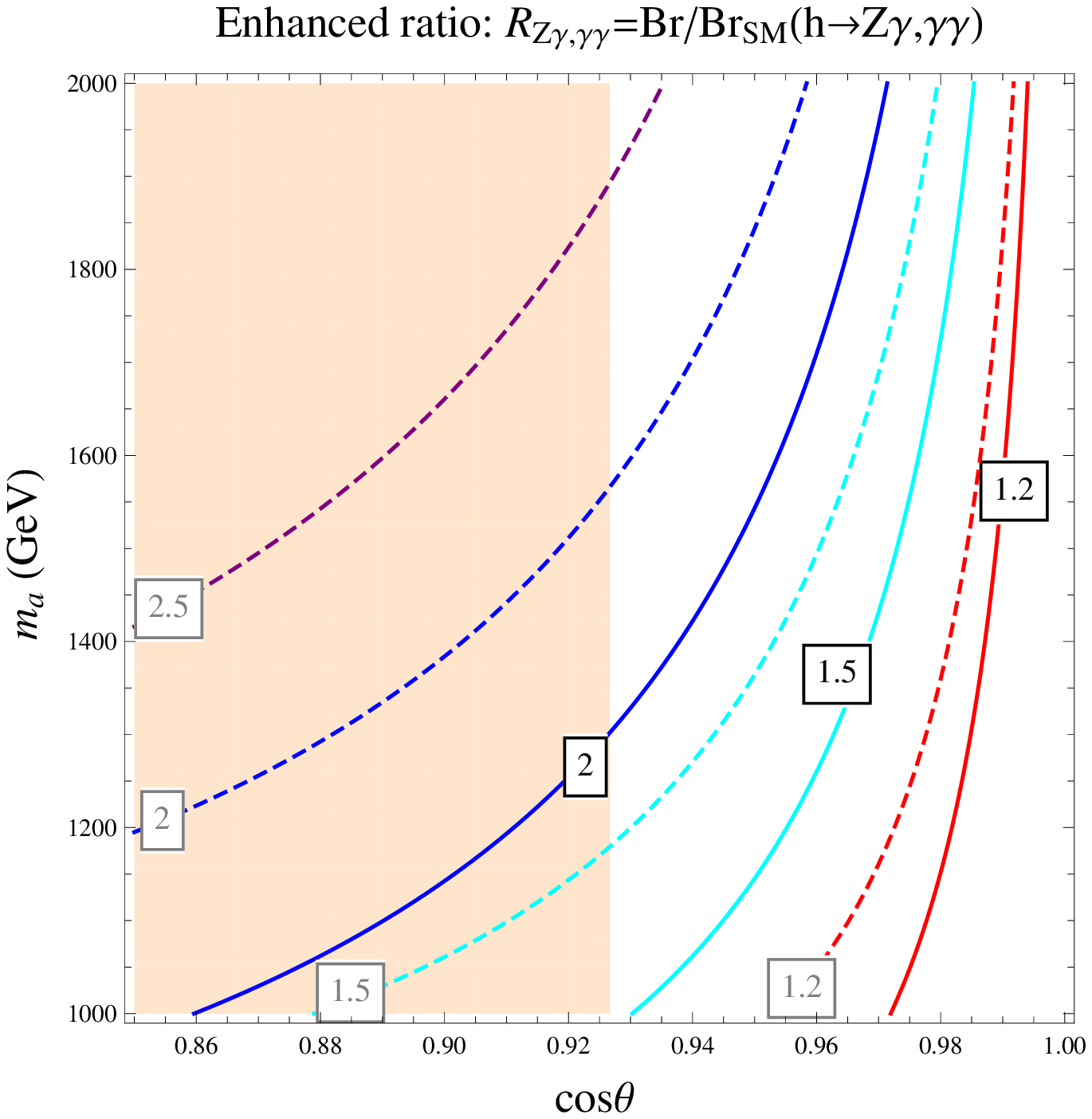}
\includegraphics[angle=0,clip,width=7.5cm]{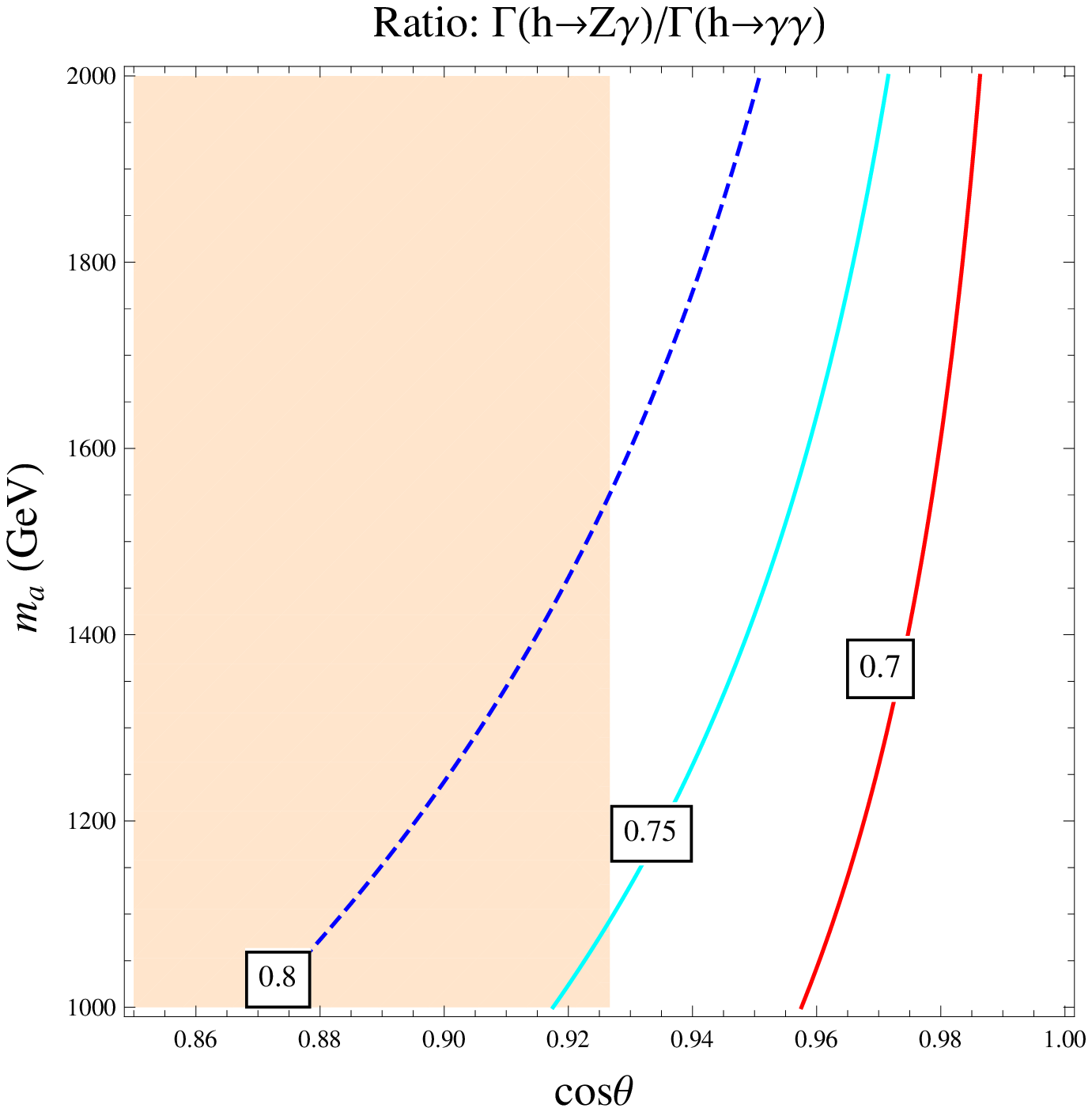}
\caption{{\small  Left Contour:  signal strengths for $Z \gamma$ mode (solid line) and $\gamma \gamma$ mode (dashed line) in the $(m_a, \cos \theta )$ plane. The purple dashed line shows the upper bound measured by  ATLAS at $68 \%$ confidence level.  Right Contour: the decay width ratio between  $Z \gamma $ mode and $ \gamma \gamma$ mode in the $(m_a, \cos \theta )$ plane. The condition  $\Delta = 0.75$  is imposed.  The light orange region is excluded by  the  EWPT requirement  $S < 0.13$. }}
\label{ratio1}
\end{center}
\end{figure}

Now we proceed to analyze  the  process of  $ h^0 \to Z \gamma$ and compare it with the experimentally better measured process $h^0 \to \gamma \gamma$. In terms of the form factor $c_{Z \gamma}$ and the rescaled branching ratio for each Higgs decay mode,  it is convenient to define the signal strength $R_{Z \gamma}$, which is one of the measurable  properties at the LHC.  Let us solely consider the gluon fusion process, where the ratio for the production cross sections is $\sigma/  \sigma_{sm} =c_t^2$, i.e. the one in composite Higgs model divided by its SM expectation. Then when assuming $c_t =1$, we have:
\beq
 R_{Z \gamma} = \frac{|c_{Z \gamma} /c_{Z \gamma}^{sm}|^2}{a_W^2 Br_{sm}^{(W W^*)} + a_Z^2 Br_{sm}^{(Z Z^*)}+|c_{\gamma \gamma} /c_{\gamma \gamma}^{sm}|^2 Br_{sm}^{(\gamma \gamma)} +|c_{Z\gamma} /c_{Z \gamma}^{sm}|^2 Br_{sm}^{(Z \gamma)} +  \cdots}\, ,
\eeq
and for the diphton process,  the signal strength $R_{\gamma \gamma}$  is calculated in analogous way by substituting $|c_{Z \gamma}/c_{Z \gamma}^{sm}|^2$ with $|c_{\gamma \gamma}/c_{\gamma\gamma}^{sm}|^2$ in the numerator, where the analytic expression for $c_{\gamma \gamma }$ in this model has already been given in \cite{cai}.
Since $ h^0 \to  Z \gamma$ and $ h^0 \to \gamma \gamma$ are
correlated to each other due to the EW symmetry breaking,
following the proposal  in~\cite{Chiang},  we define a useful parameter,  the decay
width ratio between these two modes,  to measure the degree of
correlation :
\beq
\frac{\Gamma (h^0 \to Z \gamma) }{\Gamma (h^0 \to  \gamma \gamma) }=  2  \cdot \bigg|\frac{c_{Z \gamma} }{c_{\gamma \gamma} }\bigg|^2   \frac{ (m_h^2 - m_Z^2)^3 }{ m_h^6} \, .
\eeq

Contours in Fig. [\ref{ratio}] quantitatively show behaviors of  signal strengths $R_{Z \gamma}$, $R_{\gamma \gamma}$ and
the decay width ratio $\Frac{\Gamma (h^0 \to Z \gamma) }{\Gamma (h^0 \to  \gamma \gamma) }$ in the plane of $(m_a, \cos \theta)$,
 with other parameters fixed: $g_\rho = g_a = 3.0 $ and  $ \Delta = 0.5 $.  As we can see,
 the stringent EWPT constraint comes from the $S$ parameter.  Obviously the one dimension $68 \%$ C.L.  bound $S < 0.13 $
 translates to be a condition $ \cos \theta > 0.96 $.  The left contour shows that a  signal strength as much as  $ R_{Z \gamma} \approx 1.5$  is possible to be achieved in a sizable portion of parameter region
 when we demand $ m_a > 1.2  ~\mbox{TeV}$ and $\cos \theta >0.96$.
For comparison it is shown with the dashed lines that $R_{\gamma \gamma}$ is slightly larger for a given point at the $(m_a, \cos \theta)$ plane.  On experimental side,  the most recent measurement at ATLAS  reported  an enhancement in diphoton signal strength to be $1.8^{ + 0.71}_{ - 0.59} $ with the errors linearly added ~\cite{atlas}. Therefore  the  upper bound  $ R_{\gamma \gamma} \sim  2.5$ at least  provides certain constraint on the parameter space as indicated  by  the dashed purple line in the left contour. Notice that since in our model $m_a$ is not correlated to  $ \cos \theta \equiv \cos(v/ f)$, the decoupling of spin-$1$ resonance will only happen as  $f$ is pushed into $\infty $ (i.e. $\cos \theta $ goes into $1$), and in such a limit the Standard Model is recovered. On the other hand, the right contour illustrates the correlation between $Z \gamma $ mode and $\gamma \gamma$ mode.  The value for this ratio in the SM is $\Frac{\Gamma_ {Z \gamma, sm}}{\Gamma_{\gamma \gamma, sm}} = 0.636$. Positive deviation from this value roughly indicates that $R_{Z \gamma}$ would be larger in contrast to $R_{\gamma \gamma}$, while negative deviation implies the other situation. Under the condition $\Delta =0.5$, the ratio is generally smaller than its SM value, and when we decrease $\cos \theta$ and increase $m_a$, the ratio will be further reduced. This verifies the exact trend that we observe from the left contour.

In Fig. [\ref{ratio1}],  we choose another benchmark point $g_\rho
= g_a =3.0$ and $ \Delta = 0.75 $ and draw the contours in the
same manner.  In this case,  when we demand $ S < 0.13$, the
lowest allowed value of $\cos \theta $ is tuned to be $0.92$ such
that we get more EWPT applicable parameter space. This gain is
achieved with the deduction in the enhancement rates for $h^0
\to Z \gamma, \gamma \gamma$. However more enhancement is
found for the $Z \gamma $ mode rather than for the $\gamma
\gamma$ mode,  compared with the former case.
 An inverse pattern is similarly displayed in the correlation contour, i.e. with
smaller $\cos \theta$ and bigger $m_a$, the ratio will be instead
increased. Moreover the ratio tends to be larger than in the SM
provided that the value of $\Delta $ is close  enough to $1.0$. 

\section{Conclusion} 
In this paper we discuss the circumstance that axial resonances
and vector resonances are present in the low energy particle
spectrum of composite Higgs model.  As we know that in a general
composite Higgs model,  we need vector resonance to restore the
perturbative  unitarity till  the effective  cut off scale. However
 the  solely including of  vector resonance on the other hand increases
the $S$ parameter and potentially renders it to be too big.  In the scenario that  composite Higgs emerging as a pNGB  from a spontaneously  broken  global symmetry, a
custodial symmetry is usually imposed to protect the $T$
parameter, whereas the deviation in the $S$ parameter is measured
by  $v^2 / f^2 $ without being suppressed by one loop factor,
therefore it would cause a  tension with the electroweak precision
measurement since we demand the composite scale $4 \pi f \sim
\mathcal{O}  (10~\mbox{TeV} )$  in order to reduce the amount of
fine tuning.  One solution to cure this problem is to introduce
axial resonance as it will partially relieve the stringent
constraint and pull the $S$ parameter back to the origin point.

One crucial feature in  a Composite Higgs Model  is the partial compositeness of gauge bosons.  The nonlinear realization modifies the Higgs couplings via the strong dynamics, with the consequence that the correction may not be too much small. We calculate the signal strength for $h^0 \to Z \gamma$ in the context of a truncated effective theory,  especially including the non-diagonal gauge contribution for accuracy  and we find out that in most  EWPT allowed  region,  the signal strength for $Z \gamma$ could be enhanced as much as $50 \%$. It eventually turns out that the correlation between the $Z \gamma$ and $\gamma \gamma$ modes in our  model is relatively similar to the SM expectation. Furthermore vector and axial resonances lead to simultanous enhancement for  both signals. Nevertheless the result should be testable in  future LHC experiments  and  precise measurement of  $H$-$Z$-$\gamma$ vertex  would help to  distinguish  new physics scenarios.

%\newpage
\section*{Acknowledgments}
H.Cai is  supported by the postdoc foundation
under the Grant No. 2012M510001.

\newpage
\appendix
\section{Sigma Field in  Unitary Gauge}
Here we collect all the $SO(5)$ generators such that we can calculate the sigma matrix explicitly.
\beq
T_{ij}^{\hat a} &=& \frac{i}{{\sqrt 2 }}\left[ {\delta _i^{\hat a}\delta _j^5 - \delta _j^{\hat a}\delta _i^5} \right]   \nonumber \\
T_{L,ij}^a &=&  - \frac{i}{2}\left[ {\frac{1}{2}{\varepsilon ^{abc}}\left( {\delta _i^b\delta _j^c - \delta _j^b\delta _i^c} \right) + \left( {\delta _i^a\delta _j^4 - \delta _j^a\delta _i^4} \right)} \right]   \nonumber \\
T_{R,ij}^a &=&  - \frac{i}{2}\left[ {\frac{1}{2}{\varepsilon ^{abc}}\left( {\delta _i^b\delta _j^c - \delta _j^b\delta _i^c} \right) - \left( {\delta _i^a\delta _j^4 - \delta _j^a\delta _i^4} \right)} \right]
\eeq
where $T^{\hat a}$ with $\hat a = 1,2 ,3,4$ are generators in the coset space $SO(5)/SO(4)$, while  $T^a_{L,R}$ (denoted as $T^{a_L, a_R}$ somewhere) with $a = 1,2,3$ are generators in the unbroken subgroup $SU(2)_L \times SU(2)_R$ respectively.  Since the coset space $SO(5)/SO(4)$ is symmetric,  an automorphism  of  the algebra will change the sign of  the broken generators: $T^{\hat a} \to - T^{\hat a}$.  However without conducting a field transformation $\pi^{\hat a} \to -\pi^{\hat a }$,  the choice of  sign for  the broken generators will not  alter  the CCWZ objects.

In the unitary gauge, since the other three pion fields are eaten by the external $W$ and $B$ gauge bosons, only the Higgs field enters into the  parameterization $\pi^{\hat a} =  (0,0,0, h^0)$, and  the sigma  matrix $U = \mbox{exp} (i \sqrt 2 \pi^{\hat a} T^{\hat a}/f)$ is represented by:
\beq
U = \left( \begin{array}{ccc}
I_{3 \times 3} & & \\
&  \cos \frac{h^0}{f}&- \sin \frac{h^0}{f}  \\
&  \sin \frac{h^0}{f} & \cos \frac{h^0}{f}
\end{array} \right)
\eeq
Thus for the CCWZ objects $ d_\mu^a $ and $E_\mu^{a_{L,R}}$, they have the following compact forms:
\beq
d_{\mu}  &=& - \frac{\sqrt 2 }{f} \partial_\mu  h^0 \, T^{\hat 4 }  + \frac{1}{\sqrt 2} \sin \left(\frac{h^0}{f} \right)  \left( g_0 \tilde{W}_\mu ^a- g_0^\prime \tilde{B}_\mu \delta^{a 3}\right) \delta^{a \hat i} \, T^{\hat i}  \,    \\
E_{\mu}
&=&  \left(g_0 \tilde{W}_\mu ^a T_L^a + g_0^\prime \tilde{B}_\mu T_R^3  \right)- \sin^2 \left(\frac{h^0}{2 f} \right)\left( g_0 \tilde{W}_\mu ^a- g_0^\prime \tilde{B}_\mu \delta^{a 3}\right) (T_L^a- T_R^a)
\eeq
Obviously they will translate into the previous notations used in Eq.[\ref{Emu}],  through the leading order chiral expansion.


\newpage

\begin{thebibliography}{99}

\bibitem{Higgs} G. Aad et al. [ATLAS Collaboration], Phys. Lett. B 716 (2012) 1 arXiv:1207.7214 [hep-ex]; S. Chatrchyan et al. [CMS Collaboration], Phys. Lett. B 716 (2012) 30 arXiv:1207.7235 [hep-ex].


\bibitem{CHM}
  R.~Contino, Y.~Nomura and A.~Pomarol,
  %``Higgs as a holographic pseudoGoldstone boson,''
  Nucl.\ Phys.\ B {\bf 671}, 148 (2003)
  [hep-ph/0306259];
  K.~Agashe, R.~Contino and A.~Pomarol,
  %``The Minimal composite Higgs model,''
  Nucl.\ Phys.\ B {\bf 719}, 165 (2005)
  [hep-ph/0412089];
C.~Anastasiou, E.~Furlan and J.~Santiago,
  %``Realistic Composite Higgs Models,''
  Phys.\ Rev.\ D {\bf 79}, 075003 (2009)
  arXiv:0901.2117 [hep-ph];
  G.~Panico and A.~Wulzer,
  JHEP {\bf 1109}, 135 (2011)
  arXiv:1106.2719 [hep-ph];
   S.~De Curtis, M.~Redi and A.~Tesi,
  JHEP {\bf 1204}, 042 (2012)
  arXiv:1110.1613 [hep-ph];

\bibitem{Gainer}
  J.~S.~Gainer, W.~-Y.~Keung, I.~Low and P.~Schwaller,
  %``Looking for a light Higgs boson in the $Z \gamma \to \ell \ell \gamma$ channel,''
  Phys.\ Rev.\ D {\bf 86}, 033010 (2012)
  arXiv:1112.1405 [hep-ph]

%\bibitem{atlas} The ATLAS Collaboration, ATLAS-CONF-2013-034.

%\bibitem{cms}  The CMS Collaboration, CMS PAS HIG-13-001.

%\bibitem{Dittmaier}
%  S.~Dittmaier, S.~Dittmaier, C.~Mariotti, G.~Passarino, R.~Tanaka, S.~Alekhin, J.~Alwall and E.~A.~Bagnaschi {\it et al.},
  %``Handbook of LHC Higgs Cross Sections: 2. Differential Distributions,''
%  arXiv:1201.3084 [hep-ph]


\bibitem{Carena}
  M.~Carena, I.~Low and C.~E.~M.~Wagner,
  %``Implications of a Modified Higgs to Diphoton Decay Width,''
  JHEP {\bf 1208}, 060 (2012)
  arXiv:1206.1082 [hep-ph];

\bibitem{Djouadi}
  A.~Djouadi, V.~Driesen, W.~Hollik and A.~Kraft,
  %``The Higgs photon - Z boson coupling revisited,''
  Eur.\ Phys.\ J.\ C {\bf 1}, 163 (1998)
  [hep-ph/9701342].

\bibitem{Chiang}
C.~-W.~Chiang and K.~Yagyu,
  %``Higgs boson decays to $\gamma\gamma$ and $Z \gamma$ in models with Higgs extensions,''
  arXiv:1207.1065 [hep-ph].


\bibitem{CCWZ}
S. R. Coleman, J. Wess and B. Zumino,  Phys. Rev. 177, 2239 (1969); C. G. Callan Jr., S. R. Coleman, J. Wess and B. Zumino,  Phys. Rev. 177, 2247 (1969).

\bibitem{Contino}
  R.~Contino, D.~Marzocca, D.~Pappadopulo and R.~Rattazzi,
  %``On the effect of resonances in composite Higgs phenomenology,''
  JHEP {\bf 1110}, 081 (2011)
  [arXiv:1109.1570 [hep-ph]].

\bibitem{Marzocca}
%General Composite Higgs Models
    D. Marzocca,   M. Serone and  J. Shu,
    JHEP 1208 (2012) 013
    [arXiv:1205.0770 [hep-ph]].


\bibitem{Csaki}
 %  Composite Higgs Sketch
    B. Bellazzini,  C. Csaki,   J. Hubisz, J.  Serra and J. Terning,
    JHEP 1211 (2012) 003
    arXiv:1205.4032 [hep-ph].

\bibitem{Barducci}
  D.~Barducci, A.~Belyaev, S.~De Curtis, S.~Moretti and G.~M.~Pruna,
  %``Exploring Drell-Yan signals from the 4D Composite Higgs Model at the LHC,''
  JHEP {\bf 1304}, 152 (2013)
  [arXiv:1210.2927 [hep-ph]].

\bibitem{peskin}
  M.~E.~Peskin and T.~Takeuchi,
  %``A New constraint on a strongly interacting Higgs sector,''
  Phys.\ Rev.\ Lett.\  {\bf 65}, 964 (1990);
%\bibitem{Peskin:1991sw}
%  M.~E.~Peskin and T.~Takeuchi,
  %``Estimation of oblique electroweak corrections,''
  Phys.\ Rev.\  D {\bf 46}, 381 (1992);

\bibitem{Barbieri}
  R.~Barbieri, A.~Pomarol, R.~Rattazzi and A.~Strumia,
  %``Electroweak symmetry breaking after LEP-1 and LEP-2,''
  Nucl.\ Phys.\ B {\bf 703}, 127 (2004)
  [hep-ph/0405040].

\bibitem{Barbieri:2007}
  R.~Barbieri, B.~Bellazzini, V.~S.~Rychkov and A.~Varagnolo,
  %``The Higgs boson from an extended symmetry,''
  Phys.\ Rev.\ D {\bf 76}, 115008 (2007)
  [arXiv:0706.0432 [hep-ph]].

\bibitem{pich}
  A.~Pich, I.~Rosell and J.~J.~Sanz-Cillero,
  %``Viability of strongly-coupled scenarios with a light Higgs-like boson,''
  arXiv:1212.6769 [hep-ph].

\bibitem{Lavoura}
 L. Lavoura and J. P. Silva,  %"The Oblique corrections from vector - like singlet and doublet quarks" ,
 Phys. Rev. D 47 (1993) 2046;
  H.~Cai,
%``Mono Vector-Quark Production at the LHC,''
  JHEP {\bf 1302}, 104 (2013)
  arXiv:1210.5200 [hep-ph].


\bibitem{Cacciapaglia}
  G.~Cacciapaglia, A.~Deandrea, G.~D.~La Rochelle and J.~-B.~Flament,
  %``Higgs couplings beyond the Standard Model,''
  arXiv:1210.8120 [hep-ph].

\bibitem{cai}
H. Cai,
% "Higgs decay into a diphoton in Composite Higgs Model,"
Phys. \ Rev.\ D {\bf 88}, 035018 (2013)
arXiv:1303.3833 [hep-ph].


\bibitem{tHooft}
  G.~'t Hooft and M.~J.~G.~Veltman,
  %``Scalar One Loop Integrals,''
  Nucl.\ Phys.\ B {\bf 153}, 365 (1979);
  G.~Passarino and M.~J.~G.~Veltman,
  %``One Loop Corrections for e+ e- Annihilation Into mu+ mu- in the Weinberg Model,''
  Nucl.\ Phys.\ B {\bf 160}, 151 (1979);
  A.~Denner,
  %``Techniques for calculation of electroweak radiative corrections at the one loop level and results for W physics at LEP-200,''
  Fortsch.\ Phys.\  {\bf 41}, 307 (1993) arXiv:0709.1075 [hep-ph]

\bibitem{hza}
F. Wilczek, Phys. Rev. Lett. 39 (1977) 1304;
R.N. Cahn, M.S. Chanowitz and N. Fleishon, Phys. Lett. B82 (1979) 113;
L. Bergstrom and G. Hulth, Nucl. Phys. B259 (1985) 137.

\bibitem{atlas} The ATLAS Collaboration, ATLAS-CONF-2012-168.




\end{thebibliography}
\end{document}